\documentclass[aip,jcp,reprint,%eqsecnum,
superscriptaddress,twocolumn,longbibliography,floatfix]{revtex4-2}

\usepackage{empheq}
\usepackage{graphicx}% Include figure files
\usepackage{dcolumn}% Align table columns on decimal point
\usepackage{bm}% bold math
\usepackage[utf8]{inputenc}
\usepackage[T1]{fontenc}
\usepackage{mathptmx}
\usepackage{etoolbox}
\usepackage{xcolor}
\usepackage{mathpazo}
\usepackage{bm}
\usepackage[unicode=true,
 bookmarks=true,bookmarksnumbered=false,bookmarksopen=false,
 breaklinks=false,pdfborder={0 0 0},pdfborderstyle={},backref=false,colorlinks=true]
 {hyperref}
\hypersetup{pdfborderstyle={},pdfborderstyle={},pdfborderstyle={},pdfborderstyle={},pdfborderstyle={},pdfborderstyle={},linkcolor=blue,citecolor=blue}
\usepackage{breakurl}

\usepackage{natbib}
\usepackage{amssymb,amsmath}
\usepackage{comment}

\usepackage{float}
\newcommand{\phib}{\phi_{\text{mixt}}}
\newcommand{\phim}{\phi_{\text{mono}}}
\newcommand{\Zb}{Z_{\text{mixt}}}
\newcommand{\Zm}{Z_{\text{mono}}}
\newcommand{\xL}{x_{\text{L}}}
\newcommand{\cL}{c_{\text{L}}}
\newcommand{\xS}{x_{\text{S}}}
\newcommand{\q}{q}

\newcommand\beq{\begin{equation}}
	\newcommand\eeq{\end{equation}}
\newcommand\beqa{\begin{eqnarray}}
	\newcommand\eeqa{\end{eqnarray}}
\newcommand{\nn}{\nonumber\\}
\def\bal#1\eal{\begin{align}#1\end{align}}

\makeatletter
\def\@email#1#2{%
 \endgroup
 \patchcmd{\titleblock@produce}
  {\frontmatter@RRAPformat}
  {\frontmatter@RRAPformat{\produce@RRAP{*#1\href{mailto:#2}{#2}}}\frontmatter@RRAPformat}
  {}{}
}%
\makeatother
\begin{document}

\title[]{
Predicting random close packing of binary hard-disk mixtures via third-virial-based parameters}
% Force line breaks with \\
\author{Andr\'es Santos}%
 \affiliation{
Departamento de F\'isica and Instituto de Computaci\'on Cient\'ifica Avanzada (ICCAEx), Universidad de Extremadura, E-06006 Badajoz, Spain%\\This line break forced with \textbackslash\textbackslash
}%
\author{Mariano L\'opez de Haro}
\affiliation{Instituto de Energ\'{\i }as Renovables, U.N.A.M.,
Privada Xochicalco s/n, Col. Centro, Temixco, Mor.\ 62580, Mexico}
\email{malopez@unam.mx}
\email{andres@unex.es}

\date{\today}% It is always \today, today,

\begin{abstract}
We propose a simple and accurate approach to estimate the random close packing (RCP) fraction of binary hard-disk mixtures. By introducing a parameter based on the mixture's reduced third virial coefficient---which effectively captures three-body correlations and excluded-area constraints---we show that the RCP fraction depends nearly linearly on this parameter, leading to a near-universal collapse of simulation data over a wide range of size ratios and compositions. Comparisons with previous models by Brouwers and Zaccone indicate that the present approach provides more accurate and consistent predictions. The method can be naturally extended to polydisperse mixtures with continuous size distributions and is structurally consistent with the surplus equation-of-state formulation, offering a compact framework for understanding the near universality of RCP in hard-disk systems.
\end{abstract}

\maketitle

\section{Introduction}
\label{sec1}

Ever since the introduction of the concept by Bernal,\cite{B59,B60,BM60,TS10,F13b} random close packing (RCP) has remained a subject of sustained interest across mathematics,
physics, and engineering.%
\cite{SDMK64,YCW65,SK69,F70,KFT71,VM72,QT74,S77,S80,S80b,AFLP82,B83,BT84,BGOT86,OTBDP86,HFJ90,LS90,BL92,YS93d,FSS93,W98d,TTD00,OH02,KTS02,DTSC04,UST04,%
OO04,XBOH05,KL07,DW09,FG09,BCPZ09,MSJWM10,KWP10,XR11,W12,NSD13,AST14,DW14,BT14,MB14,MLL14,SYHOO14,M15b,KVTT16,PSG17,ZT17,MB17,T18,KBS19,TJ20,LHCLL20,B21,WGLC21,B23,%
ACZMMZ23,SZKG23,MGdFW23,GBcW24,MPHFKL24,BGSO24,B24,B25,WMH25,F25,N25,Z25,Z25b,BGVO25,B25b,SWJ25,B26}
This enduring attention stems from the ubiquitous appearance of densely packed disordered structures in a wide variety of systems, including jamming transitions,\cite{TS10,OH02,DTSC04,UST04,MSJWM10,KWP10,XR11,AST14,KVTT16,PSG17,ZT17,MB17,TJ20,BGSO24} colloids,\cite{KFT71,QT74,S77} granular media,\cite{S80,MB14,M15b} glasses,\cite{NSD13} amorphous solids,\cite{BT84} ceramics,\cite{LHCLL20} and even living matter,\cite{SWJ25} as illustrated by the extensive body of work cited above.

A commonly adopted definition of RCP, although sometimes challenged as ill-defined,\cite{TTD00,KL07} identifies it as the highest achievable density at which a large collection of particles can be packed without long-range order.\cite{B83,KL07,T18}
The concept was originally formulated for monodisperse hard-sphere (HS) systems and later extended to monodisperse hard-disk (HD) systems, both of which have played a paradigmatic role in the study of packing problems. An additional complication arises from the fact that different packing protocols, whether numerical or experimental, may lead to different maximal densities.\cite{TTD00,BCPZ09}

Polydispersity further affects the RCP fraction, which helps explain why studies of polydisperse HS and HD systems are comparatively scarce relative to their monodisperse counterparts. From a theoretical standpoint, the analytic determination of the RCP fraction in such systems remains essentially an open problem. As a result, recent theoretical efforts have focused on approximate, physically motivated descriptions aimed at capturing the dominant effects of size disparity in simple mixtures.

In two recent papers,\cite{B24,B25} Brouwers proposed, using a geometric approach, the following approximation for the RCP fraction ($\phib$) of a binary HD mixture with small or moderate dispersity:
\begin{subequations}
\label{1-2}
\beq
\label{1}
\phib=\frac{\phim}{1-\mu_B(1-\phim)},\quad \mu_B\equiv\frac{(\q^2-1)^2}{2(\q^2+1)}\frac{\xL\xS}{m_2},
\eeq
\beq
\label{2}
\frac{\phib}{1-\phib}=\lambda_B\frac{\phim}{1-\phim},\quad \lambda_B\equiv \frac{1}{1-\mu_B}.
\eeq
\end{subequations}
Here, $\phim$ denotes the RCP fraction of the monodisperse HD system, $\xL$ and $\xS=1-\xL$ are the mole fractions of large and small disks, respectively, $\q\equiv\sigma_{\text{L}}/\sigma_{\text{S}}\geq 1$ is the size ratio, and $m_n=M_n/M_1^n$ are reduced moments, with $M_n=\xS\sigma_{\text{S}}^n+\xL\sigma_{\text{L}}^n$ being the moments of the size distribution.

By fitting Eq.~\eqref{1} to Monte Carlo simulation data for $\q=1.4$,\cite{note_12_2025} Brouwers obtained $\phim=0.8425$, a value slightly below the commonly accepted estimate $\phim=0.844$.\cite{BCPZ09}

\begin{figure}
           \includegraphics[width=0.85\columnwidth]{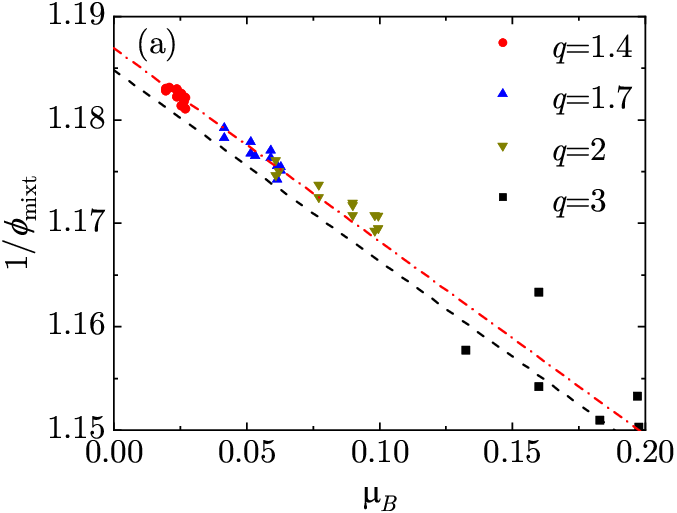}\\ \includegraphics[width=0.85\columnwidth]{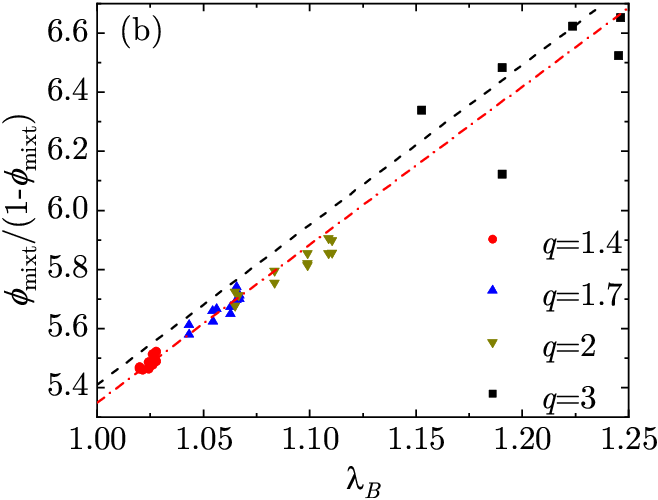}
\caption{(a) $1/\phib$ vs the parameter $\mu_B$ [Eq.~\eqref{1}] and (b) $\phib/(1-\phib)$ vs $\lambda_B$ [Eq.~\eqref{2}] for binary HD mixtures with size ratios $\q=1.4$, $1.7$, $2$, and $3$. Symbols are simulation data,\cite{note_12_2025} while the dashed and dashed-dotted lines correspond to Eqs.~\eqref{1} and \eqref{2} with $\phim=0.844$ and $\phim=0.8425$, respectively.
\label{fig1}}
\end{figure}

Two noteworthy consequences follow from Eq.~\eqref{1-2}. First, the entire dependence of $\phib$ on $\xL$ and $\q$ is encoded in the single parameter $\mu_B$ (or, equivalently, in $\lambda_B$). Consequently, two distinct binary mixtures sharing the same value of $\mu_B$ are predicted to have the same RCP fraction. Second, the inverse RCP fraction $1/\phib$ depends linearly on $\mu_B$, while the ratio $\phib/(1-\phib)$ depends linearly on $\lambda_B$. As shown in Appendix \ref{appA}, these two linear dependencies uniquely fix the functional form of Eq.~\eqref{1-2}. This implies that plots of $1/\phib$ vs $\mu_B$ and $\phib/(1-\phib)$ vs $\lambda_B$ for different mixtures should exhibit data collapse.

However, as illustrated in Fig.~\ref{fig1}, the collapse of simulation data corresponding to different mixtures is imperfect, particularly as the size ratio $\q$ increases.

The main purpose of this short paper is to demonstrate that a significantly better collapse is achieved by slightly modifying Eq.~\eqref{1-2}  and introducing an alternative ``universality'' parameter based on the third virial coefficient of the mixture. We further demonstrate that this approach can be naturally extended to polydisperse mixtures, providing a universal and practically useful framework for predicting RCP fractions in both discrete and continuous size distributions.

\begin{figure}
 \includegraphics[width=0.85\columnwidth]{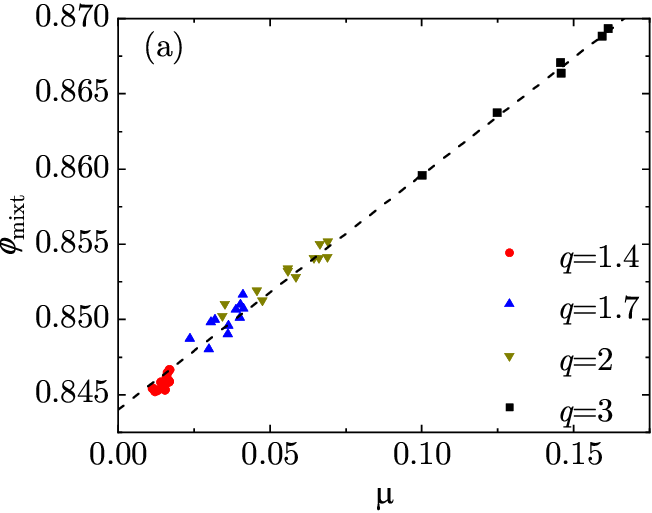}\\ \includegraphics[width=0.85\columnwidth]{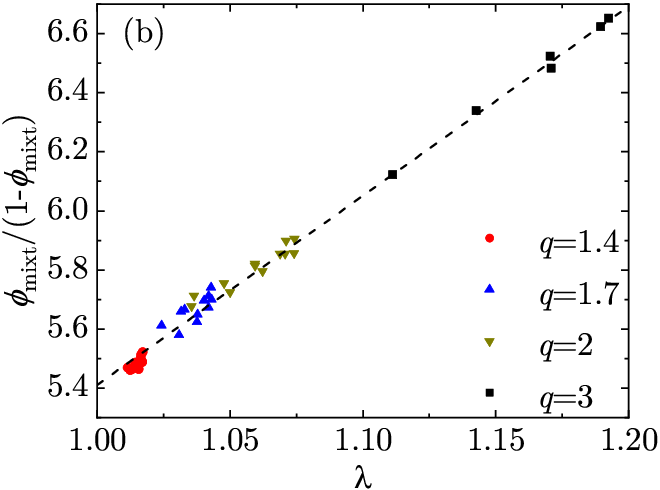}
\caption{(a) $\phib$ vs the parameter $\mu$ [Eq.~\eqref{1bis}] and (b) $\phib/(1-\phib)$ vs $\lambda$ [Eq.~\eqref{2bis}] for binary HD mixtures with size ratios $\q=1.4$, $1.7$, $2$, and $3$. Symbols are simulation data,\cite{note_12_2025} and the dashed lines correspond to our proposal [Eq.~\eqref{26_2D}] with $\phim=0.844$.
\label{fig2}}
\end{figure}

\begin{figure}
      \includegraphics[width=0.85\columnwidth]{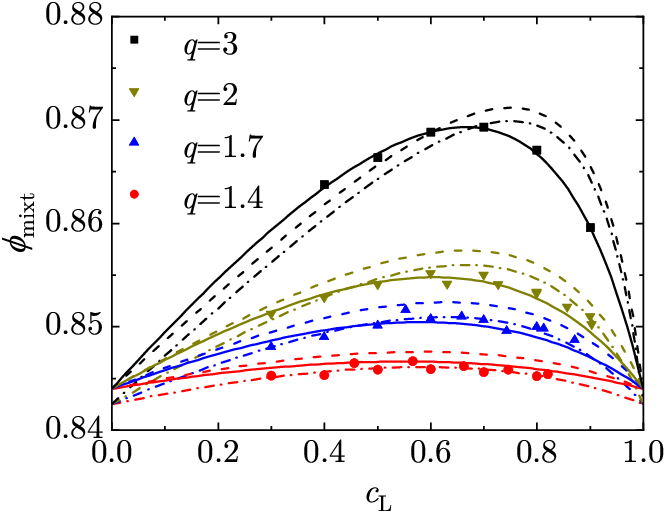}
  \caption{$\phib$ as a function of the area fraction of large disks, $\cL$, for binary HD mixtures with size ratios $\q=1.4$, $1.7$, $2$, and $3$. Symbols are simulation data.\cite{note_12_2025} Solid lines show our predictions [Eq.~\eqref{1bis}] with $\phim=0.844$, while the dashed-dotted and dashed lines correspond to Eq.~\eqref{1} with $\phim=0.8425$ and Eq.~\eqref{8} with $\phim=0.844$, respectively.
  \label{fig3}}
\end{figure}

\section{Our proposal}
\label{sec2}
In our approach, we replace Eq.~\eqref{1-2} with the following expressions:
\begin{subequations}
\label{26_2D}
\beq
\label{1bis}
\phib={\phim}+\mu(1-\phim),\quad \mu\equiv\frac{b_3-1-\left(\bar{B}_3-1\right)m_2}{b_3-3},
\eeq
\beq
\label{2bis}
\frac{\phib}{1-\phib}=\frac{\lambda}{1-\phim}-1,\quad\lambda\equiv \frac{1}{1-\mu},
\eeq
\end{subequations}
where $b_3=4\left(\frac{4}{3}-\frac{\sqrt{3}}{\pi}\right)\simeq 3.128\thinspace02$, and
\bal
\label{B3_bin}
\bar{B}_3=&\frac{\sigma_{\text{S}}^4}{M_2^2}\left[\xS^3 b_3+\frac{\xS^2 \xL}{\pi}F(q)+\frac{\xS \xL^2}{\pi}q^4F(q^{-1})+\xL^3 \q^4 b_3\right]
\eal
is the  reduced third virial coefficient of the binary mixture,\cite{S16} with
\bal
F(q)\equiv &(1+\q)^4\cos^{-1}\frac{\q^2+2\q-1}{(1+\q)^2}+8(1+\q)^2\cos^{-1}\frac{1}{1+\q}\nn
&-2\sqrt{\q(2+\q)}(3+2\q+\q^2).
\eal
The parameters $\mu$ and $\lambda$ defined in Eq.~\eqref{26_2D} are motivated by the ``surplus'' approximation for the equation of state of $d$-dimensional fluid mixtures,\cite{SYHO17,SYHOO14,S16,HSY20,SYH20} as detailed in Appendix \ref{appB}.

The key physical idea underlying Eq.~\eqref{26_2D} is that three-body correlations, encoded in the reduced third virial coefficient $\bar{B}_3$, capture the dominant effects of size disparity on dense disordered packing. In this picture, local triplet constraints and excluded-area effects largely determine how efficiently disks can arrange at jamming. Related viewpoints have recently been employed\cite{TJ20} to estimate the RCP density of three-dimensional monodisperse spherical and nonspherical particles by identifying it with the singular point obtained from the analytical continuation of the liquid-branch equation of state.

According to Eq.~\eqref{26_2D}, $\phib$ depends linearly on $\mu$, while $\phib/(1-\phib)$ depends linearly on $\lambda$. As discussed in  Appendix \ref{appA}, these two assumptions again uniquely determine the structure of Eq.~\eqref{26_2D}. Figure~\ref{fig2} shows that these simple linear relationships are well satisfied with the standard value $\phim=0.844$, particularly for $\q=3$, and that the collapse of data from different mixtures is significantly improved compared with Brouwers' model.

A complementary proposal for HD mixtures has recently been put forward by Zaccone.\cite{Z25,Z25b} In this approach, the RCP fraction is determined by equating the compressibility factors of the mixture and the monodisperse system, $\Zb(\phib)=\Zm(\phim)$. Using the Scaled Particle Theory expressions,\cite{RFL59,HFL61,LHP65} this leads to the explicit formula:\cite{Z25b}
\bal
\label{8}
\phib=&1-\frac{(1-\phim)^2}{2}\Bigg[1-m_2^{-1}\nn
&+\sqrt{(1-m_2^{-1})^2+\frac{4m_2^{-1}}{(1-\phim)^2}}\Bigg].
\eal

Figure~\ref{fig3} compares simulation data with our proposal, Brouwers' original model, and Zaccone's approach by plotting $\phib$ as a function of the area fraction of large disks,
\beq
\cL=\frac{\xL \q^2}{\xL \q^2 + \xS},
\eeq
for $\q=1.4$, $1.7$, $2$, and $3$. As observed, Brouwers' proposal [Eq.~\eqref{1}] with the fitted value $\phim=0.8425$ is accurate for $\q=1.4$ and reasonably good for $\q=1.7$, by construction, but it fails for $\q=2$ and $\q=3$. Zaccone's approach [Eq.~\eqref{8}] with $\phim=0.844$ performs fairly well for $\q=1.4$, but it underestimates $\phib$ for larger size ratios. Using the monodisperse value $\phim\simeq 0.886$\cite{Z22} in Eq.~\eqref{8}, as done in Refs.~\onlinecite{Z25,Z25b}, leads to even poorer estimates. In contrast, our third-virial-based proposal provides the best agreement across all size ratios, demonstrating its effectiveness in capturing the universality of RCP in binary HD mixtures.

\section{Extension to polydisperse mixtures}
\label{sec3}

\begin{figure}
      \includegraphics[width=0.85\columnwidth]{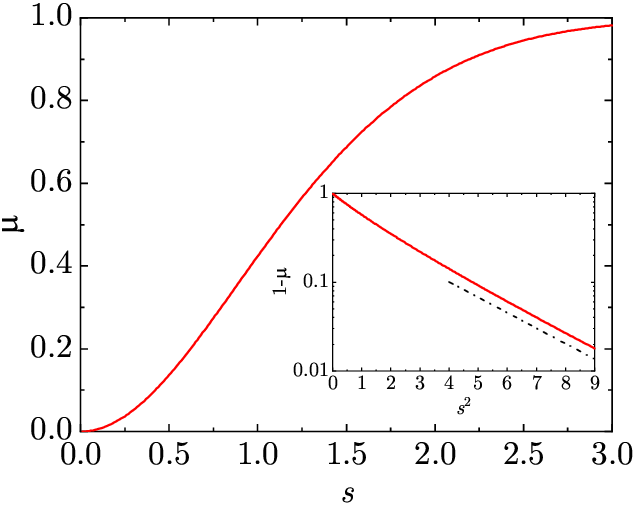}
            \caption{Plot of the parameter $\mu$ as a function of the dispersity parameter $s$ for a polydisperse mixture with a log-normal size distribution. The inset shows a logarithmic plot of $1-\mu$ as a function of $s^2$. The dashed-dotted line has a slope of $-0.4$.
  \label{fig4}}
\end{figure}

We conjecture that Eq.~\eqref{1bis} remains a good candidate for estimating the RCP fraction of a polydisperse HD mixture characterized by a continuous size distribution $f(\sigma)$. In this case, the moments of the distribution are defined as $M_n=\int_0^\infty d\sigma\, f(\sigma)\sigma^n$, and the reduced third virial coefficient is given by\cite{S16}
\bal
\bar{B}_3=&\frac{8}{\pi M_2^2}\int_0^\infty d\sigma_1\int_0^\infty d\sigma_2\int_0^\infty d\sigma_3 \,f(\sigma_1)f(\sigma_2)f(\sigma_3)\nonumber\\
&\times{A}(\sigma_{1},\sigma_2,\sigma_3),
\label{23b}
\eal
where ${A}(\sigma_{1},\sigma_2,\sigma_3)\equiv\sigma_{12}^2\mathcal{S}_{\sigma_{13},\sigma_{23}}(\sigma_{12})$ and
\bal
\mathcal{S}_{a,b}(r)=&a^2\cos^{-1}\frac{r^2+a^2-b^2}{2ar}
+b^2\cos^{-1}\frac{r^2+b^2-a^2}{2br}\nn
&-\frac{1}{2}\sqrt{2r^2(a^2+b^2)-(b^2-a^2)^2-r^4}
\label{27}
\eal
is the intersection area of two circles of radii $a$ and $b$ whose centers are separated by a distance $r$.
Note that ${A}(\sigma_{1},\sigma_2,\sigma_3)$ is a homogeneous function of degree four.

As an illustration, consider the log-normal distribution:
\beq
f(\sigma)=\frac{1}{\sigma\sqrt{2\pi s^2}}e^{-[\ln(\sigma/\sigma_0)]^2/2s^2},
\eeq
for which the moments are $M_n=\sigma_0^n e^{n^2 s^2/2}$ and the reduced moments are $m_n=e^{n(n-1)s^2/2}$.
By introducing the new variables
\beq
u=\frac{1}{3s}\ln\frac{\sigma_1\sigma_2\sigma_3}{\sigma_0^3},\quad y_1=\frac{1}{s}\ln\frac{\sigma_1}{\sigma_0}-u,\quad y_2=\frac{1}{s}\ln\frac{\sigma_2}{\sigma_0}-u,
\eeq
one has
\bal
\bar{B}_3
=&\frac{4\sqrt{3}e^{-4 s^2/3}}{\pi^2}
\int_{-\infty}^\infty dy_1\int_{-\infty}^\infty dy_2 \, e^{-(y_1^2+y_2^2+y_1 y_2)}\nn
&\times
{A}(e^{y_1 s},e^{y_2 s},e^{-(y_1+y_2) s}),
\label{C4}
\eal
where we have taken into account the homogeneity property of the function $A(\sigma_1,\sigma_2,\sigma_3)$ and  performed the Gaussian integral over $u$. The double integral in Eq.~\eqref{C4} must be evaluated numerically.

Figure~\ref{fig4} shows the parameter $\mu$ as a function of the dispersity parameter $s$. According to Eq.~\eqref{1bis}, the RCP fraction $\phib$ is expected to vary approximately linearly with $\mu(s)$.
It ranges from $\mu=0$ as $s\to 0$ to $\mu=1$ as $s\to\infty$.

The asymptotic behavior $\mu \to 1^{-}$ for large $s$ ensures that Eq.~\eqref{1bis} remains consistent with the physical bound $\phib \le 1$. The inset of Fig.~\ref{fig4} shows that $1-\mu$ decays approximately exponentially with $s^2$, indicating that the saturation is smooth rather than divergent. Thus, within the present third-virial-based framework, the mapping does not lead to unphysical packing fractions, even at large dispersity.

It should be stressed, however, that extremely large values of $s$ correspond to size distributions with exponentially broad tails, for which the physical interpretation of RCP itself becomes subtle. In such regimes, higher-order virial coefficients or a different functional form beyond the present linear mapping may be required to capture the packing constraints accurately. The present results suggest that the third virial coefficient captures the dominant geometric effects over a broad, but not arbitrarily extreme, range of dispersities.

The above considerations are independent of the specific choice of size distribution. In general, the values of $\phim$ and $\phib$ may depend on the packing protocol employed.\cite{TTD00,BCPZ09} Nevertheless, Eq.~\eqref{26_2D} is expected to remain applicable to arbitrary polydisperse mixtures, provided that $\phim$ and $\phib$ are determined using equivalent protocols.

\section{Conclusions}
\label{sec4}

In this work, we have proposed a simple yet effective modification of Brouwers' geometric approach for estimating the RCP fraction of binary HD mixtures. By introducing a parameter based on the mixture's reduced third virial coefficient---which effectively encodes three-body correlations and excluded-area constraints---we obtained a  linear dependence of $\phib$ on $\mu$ and of $\phib/(1-\phib)$ on $\lambda$, leading to a significantly improved collapse of simulation data across a wide range of size ratios. Comparisons with Brouwers' original model and Zaccone's proposal show that our approach provides a more accurate description of the RCP fraction for both small and moderate dispersity.

Furthermore, the method can be naturally extended to polydisperse mixtures with continuous size distributions, such as log-normal distributions, by computing the corresponding moments and third virial coefficient. The resulting parameter $\mu$ offers a compact and universal measure of dispersity, allowing $\phib$ to be predicted as a nearly linear function of $\mu$.

Although this extension to continuous polydisperse mixtures remains a conjecture, the excellent agreement with simulation results in the binary case suggests that it is worth testing. Its simplicity and predictive character may stimulate new simulations aimed at confirming or refuting the conjecture, thereby providing a clear direction for future work.

An interesting open question concerns the possible extension of the present framework to mixtures of nonspherical convex particles, such as ellipses or spherocylinders. In such systems, the third virial coefficient incorporates not only size effects but also orientational excluded-area (or excluded-volume) contributions.\cite{TVRT91,R93,W96,W01,B04,YVM05,KW24} It is, therefore, tempting to speculate that a suitably defined reduced third virial coefficient, combined with the corresponding monocomponent RCP fraction for the given particle shape, could provide a natural generalization of the present linear mapping in terms of $\mu$. Whether this remains a quantitatively robust approximation in the presence of orientational degrees of freedom is, however, an issue that deserves dedicated investigation.

More generally, the present formulation is structurally consistent with the surplus equation-of-state framework for mixtures, reinforcing the idea that low-order virial information captures the dominant geometric constraints governing dense disordered packings. Overall, our results indicate that the third-virial-based parameter $\mu$ provides a robust and practically useful framework for understanding the near-universality of RCP in HD mixtures, with potential extensions to more complex and polydisperse systems.

\begin{acknowledgments}
A.S.\ acknowledges financial support from Grant No.~PID2024-156352NB-I00 funded by MCIU/AEI/10.13039/501100011033 and by ERDF/EU, and from Grant No.~GR24022 funded by the Junta de Extremadura (Spain).

\end{acknowledgments}

\section*{AUTHOR DECLARATIONS}
\subsection*{Conflict of Interest}
The authors have no conflicts to disclose.
\subsection*{Author Contributions}
\textbf{Andr\'es Santos}:
Conceptualization (lead);
Formal analysis (equal);
Funding acquisition (lead);
Investigation (equal);
Methodology (equal);
Supervision (lead);
Writing -- original draft (equal);
Writing -- review \& editing (equal).
\textbf{Mariano L\'opez de Haro}:
Formal analysis (equal);
Investigation (equal);
Methodology (equal);
%Software (lead);
Writing -- original draft (equal);
Writing -- review \& editing (equal).

\section*{Data availability} Data supporting figures and numerical results are available from the corresponding author upon reasonable request.

\appendix

\section{Linear dependencies of $\phib/(1-\phib)$}
\label{appA}

Here we show that Eq.~\eqref{1-2} represents the only possible forms if one assumes that $1/\phib$ is a linear function of $\mu_B$ and that $\phib/(1-\phib)$ is a linear function of $\lambda_B\equiv 1/(1-\mu_B)$. Under these assumptions, one may write
\begin{subequations}
\label{A1}
\beq
\frac{1}{\phib}=\frac{1}{\phim}-A\mu_B,
\eeq
\beq
\frac{\phib}{1-\phib}=\frac{\phim}{1-\phim}+B\left(\lambda_B-1\right),
\eeq
\end{subequations}
where we have used the fact that $\mu_B=0$ in the monocomponent limit. Combining Eq.~\eqref{A1} one obtains
\beq
1-\left(\frac{1-\phim}{\phim}\right)^2\frac{B}{A}
+\left(\frac{1-\phim}{\phim}B-1\right)\mu_B=0.
\eeq
Since this identity must hold for arbitrary $\mu_B$, both the constant and linear terms must vanish independently. This yields $A=1/B=(1-\phim)/\phim$, in agreement with Eq.~\eqref{1-2}.

An analogous reasoning applies to our proposal. If one assumes that $\phib$ is a linear function of $\mu$ and that $\phib/(1-\phib)$ is a linear function of $\lambda\equiv 1/(1-\mu)$, one can write
\begin{subequations}
\label{A2}
\beq
\phib=\phim+C\mu,
\eeq
\beq
\frac{\phib}{1-\phib}=\frac{\phim}{1-\phim}+D\left(\lambda-1\right).
\eeq
\end{subequations}
Eliminating $\phib$ between Eq.~\eqref{A2} gives
\beq
1-(1-\phim)^2\frac{D}{C}
+\left[(1-\phim)D-1\right]\mu=0.
\eeq
Requiring this expression to be valid for all $\mu$ leads to $C=1/D=1-\phim$, in agreement with Eq.~\eqref{26_2D}.

\section{The surplus equation of state for mixtures}
\label{appB}

By imposing a set of consistency conditions, a fundamental-measure-theory  approach linking the equilibrium equation of state of a three-dimensional monocomponent HS fluid to that of a three-dimensional polydisperse HS mixture was derived.\cite{S12,S12c,SYHOO14,S16} In this so-called surplus approximation, the compressibility factor of the mixture, $\Zb$, is expressed in terms of that of the monocomponent fluid, $\Zm$, as
\beq
\Zb(\phib)=\alpha\frac{\phim}{\phib} \Zm(\phim)+\frac{1-\alpha/\lambda}{1-\phib},
\label{19}
\eeq
where the packing fraction $\phim$ of the monocomponent fluid is related to the packing fraction $\phib$ of the polydisperse mixture through
\begin{subequations}
\begin{equation}
\phim=\frac{\phib}{\phib+\lambda(1-\phib)},
\end{equation}
\begin{equation}
\frac{\phib}{1-\phib}=\lambda\frac{\phim}{1-\phim}.
\label{18a}
\end{equation}
\end{subequations}
The parameters $\lambda$ and $\alpha$ can be determined by imposing consistency with the (reduced) second and third virial coefficients of the mixture,\cite{S16} leading to
\begin{equation}
\lambda=\frac{\bar{B}_2-1}{b_2-1}\frac{b_3-2b_2+1}{\bar{B}_3-2\bar{B}_2+1},\quad
\alpha=\lambda^2\frac{\bar{B}_2-1}{b_2-1}.
\label{26}
\end{equation}

Although initially derived for three-dimensional HS systems,\cite{S12c,SYHOO14} Eqs.~\eqref{19}--\eqref{26} can be generalized to any dimensionality $d\neq 3$.\cite{SYHO17,HSY20} In the particular case of HD systems ($d=2$), one has $b_2=2$ and $\bar{B}_2=1+m_2^{-1}$, so that the parameter $\lambda$ reduces to the expression given in Eq.~\eqref{26_2D}.

While the surplus equation of state is restricted to equilibrium states, we borrow two structural features from Eq.~\eqref{18a} in our proposal to relate the RCP of the HD mixture ($\phib$) to that of the monocomponent system ($\phim$). First, we assume that all the details of the mixture composition are entirely encapsulated in the universality parameter $\lambda$. Second, we assume that the ratio $\phib/(1-\phib)$ is a linear function of $\lambda$. To these two assumptions, we add the condition that $\phib$ is a linear function of $\mu=1-\lambda^{-1}$. As shown in Appendix~\ref{appA}, the combination of these three conditions uniquely leads to Eq.~\eqref{26_2D}.

\section*{REFERENCES}
\bibliography{C:/AA_D/Dropbox/Mis_Dropcumentos/bib_files/liquid}
\end{document}